\documentclass[global,twocolumn,referee]{svjour}
\usepackage{amsmath}

  \usepackage{graphics}

\journalname{myjournal}
\begin{document}

\title{Polarization properties of subwavelength hole arrays consisting of rectangular holes}

\author{Xi-Feng Ren, Pei Zhang, Guo-Ping Guo\thanks{E-mail: \email:gpguo@ustc.edu.cn}, Yun-Feng Huang, Zhi-Wei Wang, Guang-Can Guo}

\institute{
  Key Laboratory of Quantum Information, University of Science and Technology of China, Hefei
230026, People's Republic of China }

\date{Received: date / Revised version: date}

\maketitle

\begin{abstract}
Influence of hole shape on extraordinary optical transmission was
investigated using hole arrays consisting of rectangular holes with
different aspect ratio. It was found that the transmission could be
tuned continuously by rotating the hole array. Further more, a phase
was generated in this process, and linear polarization states could
be changed to elliptical polarization states. This phase was
correlated with the aspect ratio of the holes. An intuitional model
was presented to explain these results.

\end{abstract}

\section{introduction}
In metal films perforated with a periodic array of subwavelength
apertures, it has long been observed that there is an unusually high
optical transmission\cite{1}. It is believed that metal surface
plays a crucial role and the phenomenon is mediated by surface
plasmon polaritons (SPPs) and there is a process of transforming
photon to SPP and back to photon\cite{4,crucial,ebbesen5}. This
phenomenon can be used in various applications, for example,
sensors, optoelectronic device,
etc\cite{williams,brolo,nahata,luo,shinada,ebbeson07}. Polarization
properties of nanohole arrays have been studied in many
works\cite{Elli04,Gordon04,Altew05}. Recently, orbital angular
momentum of photons was explored to investigate the spatial mode
properties of surface plasmon assisted transmission
\cite{ren061,ren062}. It is also showed that entanglement of photon
pairs can be preserved when they respectively travel through a hole
array \cite{ren062,Alt,energy}. Therefore, the macroscopic surface
plasmon polarizations, a collective excitation wave involving
typically $10^{10}$ free electrons propagating at the surface of
conducting matter, have a true quantum nature. However, the
increasing use of EOT requires further understanding of the
phenomenon.

The polarization of the incident light determines the mode of
excited SPP which is also related to the periodic structure. For the
manipulation of light at a subwavelength scale with periodic arrays
of holes, two ingredients exist: shape and
periodicity\cite{4,crucial,ebbesen5,Elli04,klein,Ruan,sarra}.
Influence of unsymmetrical periodicity on EOT was discussed in
\cite{renapl}. Influence of the hole shape on EOT was also observed
recently\cite{klein,sarra}, in which the authors mainly focused on
the transmission spectra. In this work, we used rectangle hole
arrays to investigate the influence of hole shape on the
polarization properties of EOT. It is found that linear polarization
states could be changed to elliptical polarization states and a
phase could be added between two eigenmode directions. The phase was
changed when the aspect ratio of the rectangle holes was varied. The
hole array was also rotated in the plane perpendicular to the
illuminate beam. The optical transmission was changed in this
process. It strongly depended on the rotation angle, in other words,
the angle between polarization of incident light and axis of hole
array, as in the case with unsymmetrical hole array
structure\cite{renapl}.

\section{experimental results and modeling}
\subsection{Relation between transmission efficiency and photon polarization}
\begin{figure}[b]
\resizebox{0.45\textwidth}{!}{%
  \includegraphics{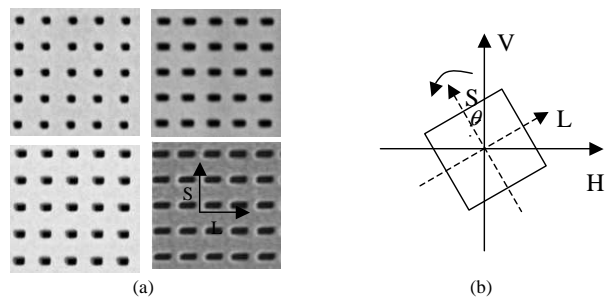}
} \caption{(Color online)The rectangle hole arrays. (a) Scanning
electron microscope pictures. (b) Rotation direction. S (L) is the
axis of short (long) edge of rectangle hole; H(V) is horizontal
(vertical) axis.}
\end{figure}
\begin{figure}[b]
\resizebox{0.45\textwidth}{!}{%
  \includegraphics{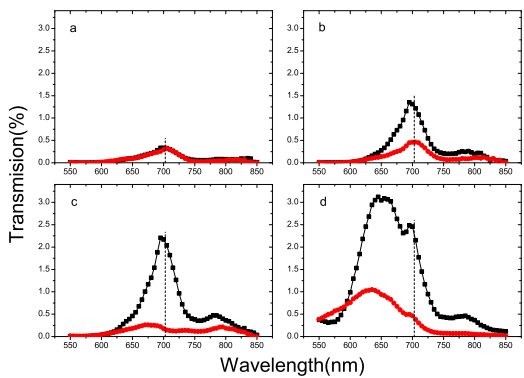}
} \caption{(Color online)Hole array transmittance as a function of
wavelength for rotation angle $\theta=0\textordmasculine$(black
square dots) and $90\textordmasculine$(red round dots)(holes for a,
b, c, and d are $100nm\times 100nm$, $100nm\times 150nm$,
$100nm\times 200nm$, and $100nm\times 300nm$ respectively). The
dashed vertical lines indicate the wavelength of $702 nm$ used in
the experiment.}
\end{figure}

Fig. 1(a) is a scanning electron microscope picture of part of our
hole arrays. The hole arrays are produced as follows: after
subsequently evaporating a $3$-$nm$ titanium bonding layer and a
$135$-$nm$ gold layer onto a $0.5$-$mm$-thick silica glass
substrate, a focused ion beam etching system is used to produce
rectangle holes ($100nm\times 100nm$, $100nm\times 150nm$,
$100nm\times 200nm$, $100nm\times 300nm$ respectively) arranged as a
square lattice ($520nm$ period). The area of the hole array is
$10\mu m\times 10\mu m$.

Transmission spectra of the hole arrays were recorded by a silicon
avalanche photodiode single photon counter couple with a
spectrograph through a fiber. White light from a stabilized
tungsten-halogen source passed though a single mode fiber and a
polarizer (only vertical polarized light can pass), then illuminated
the sample. The hole arrays were set between two lenses of $35 mm$
focal length, so that the light was normally incident on the hole
array with a cross sectional diameter about $10\mu m$ and covered
hundreds of holes. The light exiting from the hole array was
launched into the spectrograph. The hole arrays were rotated
anti-clockwise in the plane perpendicular to the illuminating light,
as shown in Fig. 1(b). Transmission spectra of the hole arrays for
rotation angle $\theta=0\textordmasculine $ and $90\textordmasculine
$ were given in Fig. 2. There were large difference between the two
cases, which was also observed in \cite{klein}.
\begin{figure}[b]
\resizebox{0.45\textwidth}{!}{%
  \includegraphics{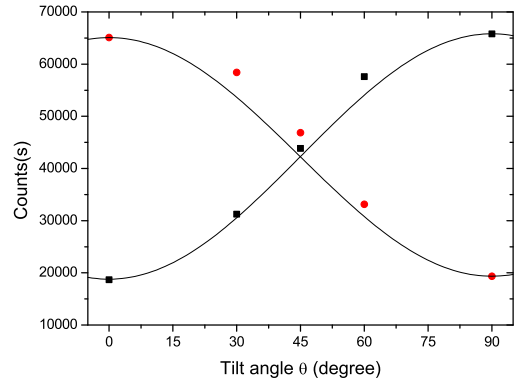}
}
\caption{(Color online)Transmittance as a function rotation angle
$\theta$ for photons in $702nm$ wavelength($100nm\times 300nm$
holes). Red round dots and black square dots are the counts for $V$
and $H$ photons respectively. The lines come from theoretical
calculation.}
\end{figure}

Further, the typical hole array($100nm\times 300nm$ holes) was
rotated anti-clockwise in the plane perpendicular to the
illuminating light(see Fig.1 (b)). Transmission efficiencies of $H$
and $V$ photons(702nm wavelength) were measured with rotation angle
$\theta=0\textordmasculine , 30\textordmasculine ,
45\textordmasculine, 60\textordmasculine$, and $90\textordmasculine
$ respectively, as shown in Fig. 3. They were varied with $\theta$.
To explain the results, we gave a simple model. For our sample,
photons with $702nm$ wavelength will excite the SPP eigenmodes
$(0,\pm 1)$ and $(\pm 1, 0)$. Since the SPPs were excited in the
directions of long (L) and short (S) edges of rectangle holes, we
suspected that this two directions were eigenmode-directions for our
sample. The polarization of illuminating light was projected into
the two eigenmode-directions to excite SPPs. After that, the two
kinds of SPPs transmitted the holes and irritated light with
different transmission efficiencies $T_{L}$ and $T_{S}$
respectively. For light whose polarization had an angle $\theta$
with the $S$ direction, the transmission efficiency $T_{\theta}$
will be

\begin{equation}
T_{\theta }=T_{S}\cos^2 (\theta)+T_{L}\sin^2 (\theta).
\end{equation}
This equation was also given in the works\cite{sarra,renapl}. Due to
the unequal values of $T_{L}$ and $T_{S}$, the whole transmission
efficiency was varied with angle $\theta$. So if we know the
transmission spectra for enginmode-directions (here L and S), we can
calculate out the transmission spectra (including the heights and
locations of peaks) for any $\theta$. The theoretical calculations
were also given in Fig. 3, which agreed well with the experimental
data. The similar results were also observed when the hole arrays
($100nm\times 150nm$ and $100nm\times 200nm$) were used. With this
model, the transmission efficiency can be continuously tuned in a
certain range.

\subsection{Influence of hole shape on photon polarization}
To investigate the polarization property of the hole array, we used
the method of polarization state tomography. Experimental setup was
shown in Fig. 4. White light from a stabilized tungsten-halogen
source passed though single mode fiber and $4 nm$ filter (center
wavelength 702 nm) to generate 702nm wavelength photons.
Polarization of input light was controlled by a polarizer, a HWP
(half wave plate, 702nm) and a QWP (quarter wave plate, 702nm). The
hole array was set between two lenses of $35 mm$ focal length.
Symmetrically, a QWP, a HWP and a polarizer were combined to analyze
the polarization of transmitted photons. For arbitrary input states,
the output states were measured in the four bases: $H$, $V$,
$1/\sqrt{2}(|H\rangle+|V\rangle)$, and
$1/\sqrt{2}(|H\rangle+i|V\rangle)$. With these experimental data, we
could get the density matrix of output states, which gave the full
polarization characters of transmitted photons. For example, in the
case of $\theta=0\textordmasculine $, for input state
$1/\sqrt{2}(|H\rangle+e^{I*0.5\pi}|V\rangle)$, four counts (8943,
31079, 3623 and 21760) were recorded when we used the four detection
bases. The density matrix was calculated as:
\begin{equation}
\left( \begin{matrix}
0.223           & -0.410 - 0.043i\\
               -0.410 + 0.043i & 0.777 \\  \end{matrix} \right),
\end{equation}
which had a fidelity of 0.997 with the pure state
$0.472|H\rangle+0.882e^{I*0.967\pi}|V\rangle$. Compared this state
with the input state, we found that not only the ratio of
$|H\rangle$ and $|V\rangle$ was changed, but also a phase
$\varphi=0.467\pi$ was added between them. The similar phenomenon
was also observed when the input state was
$1/\sqrt{2}(|H\rangle+|V\rangle)$ and in this case
$\varphi=0.442\pi$. We also considered the cases for
$\theta=30\textordmasculine , 45\textordmasculine,
60\textordmasculine$, and $90\textordmasculine $. The experimental
density matrices had the fidelities all larger than 0.960 with the
theoretical calculations, where $\varphi=(0.462\pm 0.053)\pi$. It
can be seen that the phase $\varphi$ was hardly influenced by the
rotation.

\begin{figure}[b]
\resizebox{0.45\textwidth}{!}{%
  \includegraphics{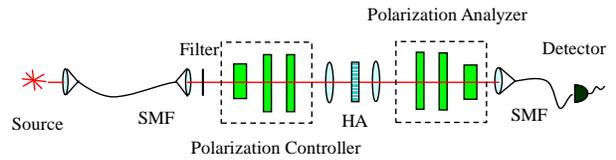}
}
\caption{(Color online)Experimental setup to investigate the
polarization property of our rectangle hole array. Polarization of
input light was controlled by a polarizer, a HWP and a QWP. The hole
array was set between two lenses of $35 mm$ focal length.
Symmetrically, a QWP, a HWP and a polarizer were combined to analyze
the polarization of transmitted photons.}
\end{figure}
\begin{figure}[b]
\resizebox{0.45\textwidth}{!}{%
  \includegraphics{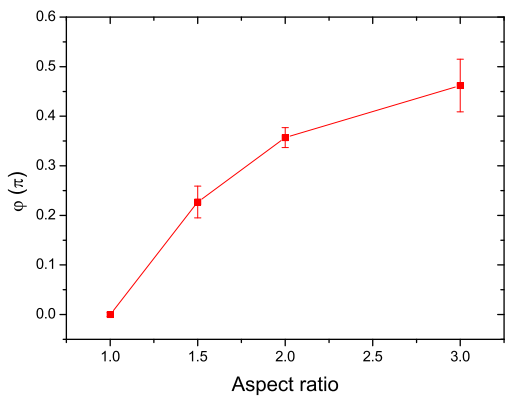}
} \caption{(Color online)Relation between birefringent phase
$\varphi$ and hole shape aspect ratio. $\varphi$ becomes lager when
the aspect ratio increases.}
\end{figure}

To study the dependence of phase $\varphi$ with the hole shape, we
performed the same measurements on other hole arrays which were
shown in Fig. 1. It was found that $\varphi$ was changed with the
aspect ratio of the rectangle holes. Fig. 5 gave the relation
between $\varphi$ and aspect ratio. The phases are $0$, $(0.227\pm
0.032)\pi$, $(0.357\pm 0.020)\pi$ and $(0.462\pm 0.053)\pi$ for
aspect ratio $1$, $1.5$, $2.0$ and $3.0$ respectively. As mentioned
above, period is another important parameter in the EOT experiments.
Since no similar result was observed for hole arrays with
symmetrical periods, a special quadrate hole array(see Fig. 1 of
\cite{renapl}) was also investigated to show the influence of the
hole period. We found that even the periods were different in two
directions, there was no birefringent phenomenon($\varphi=0$).

This birefringent phenomenon might be explained with the propagating
of SPPs on the metal surface. As we know, the interaction of the
incident light with surface plasmon is made allowed by coupling
through the grating momentum and obeys conservation of momentum
\begin{equation}
\overrightarrow{k}_{sp}=\overrightarrow{k}_{0}\pm
i\overrightarrow{G}_{x}\pm j\overrightarrow{G}_{y},
\end{equation}
where $\overrightarrow{k}_{sp}$ is the surface plasmon wave vector,
$\overrightarrow{k}_{0}$ is the component of the incident wave
vector that lies in the plane of the sample,
$\overrightarrow{G}_{x}$ and $\overrightarrow{G}_{y}$ are the
reciprocal lattice vectors, and i, j are integers. Usually,
$G_{x}=G_{y}=2\pi/d$ for a square lattice, and relation
$\overrightarrow{k}_{sp}*d=m\pi$ was satisfied, where $m$ was the
band index\cite{Teje}. While for our rectangle hole arrays, the
length of holes in $L$ direction was changed form $150nm$ to
$300nm$, which was not as same as it in $S$ direction. Though
$G_{x}=G_{y}=2\pi/d$ for our rectangle hole array, the time for
surface plasmon polariton propagating in the $L$ direction must be
influenced by the aspect ratio of hole shape, which could not be
same as that in the $S$ direction. A phase difference $\varphi$ was
generated between the two directions, leading the birefringent
phenomenon. Due to the absorption or scattering of the SPPs and
scattering at the hole edges, it is hard to give the accurate value
of the phase or the exact relation between the phase and aspect
ratio of holes. Even so, $\varphi$ could be controlled by changing
the hole shape. As a contrast, there was no birefringent phenomenon
observed when the quadrate hole array(see Fig. 1 of \cite{renapl})
was used. The reason was that phase $G_{x}*d_{x}$ always equal to
$G_{y}*d_{y}$, even $G_{x}\neq G_{y}$ for the quadrate hole array.

\section{conclusion}
In conclusion, rectangle hole array was explored to study the
influence of hole shape on EOT, especially the properties of photon
polarization. Because of the unsymmetrical of the hole shape, a
birefringent phenomenon was observed. The phase was determined by
the hole shape, which gave us a potential method to control this
birefringent process. It was also found that the transmission
efficiency can be tuned continuously by rotating the hole array.
These results might be explained using an intuitional model based on
surface plasmon eigenmodes.

This work was funded by the National Fundamental Research Program,
National Natural Science Foundation of China (10604052), Program for
New Century Excellent Talents in University, the Innovation Funds
from Chinese Academy of Sciences, the Program of the Education
Department of Anhui Province (Grant No.2006kj074A). Xi-Feng Ren also
thanks for the China Postdoctoral Science Foundation (20060400205)
and the K. C. Wong Education Foundation, Hong Kong.

\end{document}